\documentstyle[12pt]{article}
\def\Re{{\cal R \mskip-4mu \lower.1ex \hbox{\it e}\,}}
\def\Im{{\cal I \mskip-5mu \lower.1ex \hbox{\it m}\,}}
\def\ie{{\it i.e.}}

\def\etc{{\it etc}}
\def\etal{{\it et al.}}

\def\sub#1{_{\lower.25ex\hbox{$\scriptstyle#1$}}}
\def\sul#1{_{\kern-.1em#1}}
\def\sll#1{_{\kern-.2em#1}}
\def\sbl#1{_{\kern-.1em\lower.25ex\hbox{$\scriptstyle#1$}}}
\def\ssb#1{_{\lower.25ex\hbox{$\scriptscriptstyle#1$}}}
\def\sbb#1{_{\lower.4ex\hbox{$\scriptstyle#1$}}}

\def\gev{\,{\rm GeV}}

\def\to{\rightarrow}
\def\mh{\ifmmode m\sbl H \else $m\sbl H$\fi}
\def\mch{\ifmmode m_{H^\pm} \else $m_{H^\pm}$\fi}
\def\mt{\ifmmode m_t\else $m_t$\fi}
\def\mc{\ifmmode m_c\else $m_c$\fi}
\def\mz{\ifmmode M_Z\else $M_Z$\fi}
\def\mw{\ifmmode M_W\else $M_W$\fi}
\def\mws{\ifmmode M_W^2 \else $M_W^2$\fi}
\def\mhs{\ifmmode m_H^2 \else $m_H^2$\fi}
\def\mzs{\ifmmode M_Z^2 \else $M_Z^2$\fi}
\def\mts{\ifmmode m_t^2 \else $m_t^2$\fi}
\def\mcs{\ifmmode m_c^2 \else $m_c^2$\fi}
\def\mchs{\ifmmode m_{H^\pm}^2 \else $m_{H^\pm}^2$\fi}
\def\ztwo{\ifmmode Z_2\else $Z_2$\fi}
\def\zone{\ifmmode Z_1\else $Z_1$\fi}
\def\mtwo{\ifmmode M_2\else $M_2$\fi}
\def\mone{\ifmmode M_1\else $M_1$\fi}
\def\tb{\ifmmode \tan\beta \else $\tan\beta$\fi}
\def\xw{\ifmmode x\sub w\else $x\sub w$\fi}
\def\ch{\ifmmode H^\pm \else $H^\pm$\fi}
\def\lum{\ifmmode {\cal L}\else ${\cal L}$\fi}
\def\inpb{\ifmmode {\rm pb}^{-1}\else ${\rm pb}^{-1}$\fi}
\def\infb{\ifmmode {\rm fb}^{-1}\else ${\rm fb}^{-1}$\fi}
\def\epem{\ifmmode e^+e^-\else $e^+e^-$\fi}
\def\ppb{\ifmmode \bar pp\else $\bar pp$\fi}

\def\half{\textstyle{{1\over 2}}}

\newskip\zatskip \zatskip=0pt plus0pt minus0pt
\def\matth{\mathsurround=0pt}
\def\lsim{\mathrel{\mathpalette\atversim<}}
\def\gsim{\mathrel{\mathpalette\atversim>}}
\def\atversim#1#2{\lower0.7ex\vbox{\baselineskip\zatskip\lineskip\zatskip
  \lineskiplimit 0pt\ialign{$\matth#1\hfil##\hfil$\crcr#2\crcr\sim\crcr}}}

\renewcommand{\thefootnote}{\fnsymbol{footnote}}

\begin{document} \begin{titlepage}

\rightline{\vbox{\halign{&#\hfil\cr
&ANL-HEP-PR-92-102\cr
&MAD-PH-728\cr
&OTIS-499\cr
&November 1992\cr}}}
\vspace{0.5in}
\begin{center}

{\Large\bf
A Separate Higgs?}
\medskip
\vskip0.5in

\normalsize V.\ BARGER$^a$, N.G.\ DESHPANDE$^b$, J.L. HEWETT$^{b,c}$, and
T.G.\ RIZZO$^{b,c}$
\\ \smallskip
\medskip

$^a$ Department of Physics, University of Wisconsin, Madison, WI   53706\\
\smallskip
$^b$ Institute of Theoretical Science, University of Oregon, Eugene, OR 97403\\
\smallskip
$^c$ High Energy Physics Division, Argonne National Laboratory,
Argonne, IL 60439\\
\smallskip

\end{center}
\vskip1.0in

\begin{abstract}

We investigate the possibility of a multi-Higgs doublet model where the
lightest
neutral Higgs boson ($h^0$) decouples from the fermion sector.  We are
partially
motivated by the four $\ell^+\ell^-\gamma\gamma$ events with
$M_{\gamma\gamma}\simeq60$\,GeV recently observed by the L3 collaboration,
which could be a signal for $Z\to (Z^*\to \ell^+\ell^-)+(h^0\to \gamma\gamma)$.
Collider signatures for the additional physical Higgs bosons present in such
models are discussed.

\end{abstract}

\renewcommand{\thefootnote}{\arabic{footnote}} \end{titlepage}


Although the predictions of the Standard Model (SM) are in excellent agreement
with data\cite{sm}, the symmetry breaking mechanism responsible for generating
fermion and gauge boson masses remains mysterious.   In the SM, a single
scalar doublet is the source of spontaneous symmetry breaking via
the Higgs mechanism\cite{hhg}.  However, the possibility of an enlarged Higgs
sector beyond the minimal one-doublet version is consistent with data, and is
naturally present in many theories which go beyond the SM.
It is also possible that the symmetry breaking mechanism responsible for
giving masses to gauge bosons is separate from that which generates the
fermion masses \cite{them}.  An example  is technicolor models,
where fermion masses arise from extended technicolor.  Since experimental
searches for the remnants of spontaneous symmetry breaking depend
sensitively on the nature of the scalar sector, all
model possibilities must be explored.  In this context, we examine a
scenario wherein the
lightest CP-even neutral Higgs field ($h^0$) does not couple to fermions, but
has essentially SM-like couplings to the $W$ and $Z$ gauge bosons.  If such
a particle were sufficiently light, it would be produced at LEP via the
reaction
$\epem\to Z^*h^0$ and an excess of $\ell^+\ell^-\gamma\gamma$ or
$q\bar q\gamma\gamma$ final states could be observed since then the $2\gamma$
decay of $h^0$ would be dominant.
The L3 Collaboration has recently reported four $\ell^+\ell^-
\gamma\gamma$ events with $M_{\gamma\gamma}\simeq 60\gev$\cite{lthree} which
are not explained by SM processes.
Motivated by the L3 result, we explore
the simplest scenario in which the $h^0$ properties above are obtained.

A simple  realization is possible in a model with two Higgs doublets, one of
which primarily generates the $W$ and $Z$ boson masses and the other is
coupled to the fermions.  This is often referred to as model I in the
literature\cite{hhg}.  Denoting the multiplets by $\phi_{1,2}$, we impose
the discrete symmetry $\phi_1\leftrightarrow -\phi_1$ under which
the full Lagrangian and the fermion fields are fully invariant.  This
ensures that only $\phi_2$ couples to fermions and generates the fermion masses
and also guarantees that tree-level flavor changing neutral currents are
absent.
Demanding CP invariance for simplicity, the most general Higgs potential is
then given by\cite{hhg}
\begin{eqnarray}
V & = & \lambda_1(\phi_1^\dagger\phi_1-v_1^2)^2
+\lambda_2(\phi_2^\dagger\phi_2-
v_2^2)^2+\lambda_3[(\phi_1^\dagger\phi_1-v_1^2)+(\phi_2^\dagger\phi_2-v_2^2)]^2
\nonumber \\
&&\quad{}+\lambda_4[(\phi_1^\dagger\phi_1)(\phi_2^\dagger\phi_2)-
(\phi_i^\dagger\phi_2)
(\phi_2^\dagger\phi_1)] + \lambda_6[\Im(\phi_1^\dagger\phi_2)]^2 \,.
\end{eqnarray}
Here the Yukawa couplings $\lambda_i$ are real, and $v_{1,2}$ are the vacuum
expectation
values (VEV) of $\phi_{1,2}$ which are subject to the constraint $v_1^2+v_2^2
\equiv v^2$, where $v$ is the usual VEV of the SM.  The ratio of VEV's is
parameterized as $\tan\beta\equiv v_2/v_1$.
If we had tried to construct a
model wherein CP was spontaneously violated and the symmetry
$\phi_1 \leftrightarrow -\phi_1$ was strictly enforced we would find that these
two demands are incompatible{\cite {wine}}.
As usual, the Higgs mass spectrum will consist of a
neutral CP-odd field, $A$ ($m_A=v\sqrt{\lambda_6}$), a pair of charged
scalars, $H^\pm$ ($m_{H^\pm}=v\sqrt{\lambda_4}$), and a pair of CP-even
scalars whose mass matrix is given by Eq.~(4.15) in Ref.~2
with $\lambda_5$ now set to zero.
Diagonalization of this matrix through a rotation by an angle $\alpha$
($-\pi/2\leq\alpha\leq 0$) gives the mass eigenstate basis $\{h^0,\ H^0\}$ with
$m_{h^0} < m_{H^0}$
by definition.  In terms of $\alpha$ and $\beta$, the couplings of $h^0,\ H^0,\
A$, and $H^\pm$ to quarks and leptons are given in Ref.~2.
It is the field $h^0$ that we will decouple from
fermions while maintaining its couplings to the $W$ and $Z$ at close to SM
strength.

In order to sufficiently inhibit the $h^0f\bar f$  couplings, it is immediately
apparent from the structure of the couplings that we must restrict
$\cos\alpha/\sin\beta\ll 1$. In practice, we will require
$\alpha$ to be within $\simeq 10^{-3}$ of $-\pi/2$, which implies that
$\lambda_3v_1v_2\ll\lambda_iv_i^2\ (i=1,2)$ and results in
$m_{h^0}\simeq 2\sqrt{\lambda_1}v\cos\beta$ and $ m_{H^0}\simeq
2\sqrt{\lambda_2}
v\sin\beta$\ as well as $\lambda_2>\lambda_1/\tan^2\beta$.
There is no symmetry that will allow us to set $\lambda_3 = 0$, and therefore
the parameters in the Higgs potential must be fine-tuned to achieve
sufficient decoupling.  Even if we tune $\alpha$
to $-\pi/2$ at the tree-level and force the $h^0f\bar f$ coupling to zero,
loop diagrams involving both $W^\pm$ and $H^\pm$ contributions
will induce such couplings
proportional to $m_f$.  As in other fine tuned scenarios, we must choose the
value of $\alpha$ at the one-loop level to insure that the
$h^0 \to f\bar f$ rates remain small.

Since $h^0\to f\bar f$ is forbidden, the dominant decay of $h^0$ when the
$h^0$ lies below the $WW^*$ threshold is to the
$\gamma\gamma$ final state through the induced $h^0\gamma\gamma$ coupling
at one-loop.  In order to calculate the rate for $h^0\to\gamma\gamma$
(and for $H^0,A\to\gamma\gamma$), we need the $h^0H^+H^-$,
$H^0H^+H^-$, and $AH^+H^-$ couplings. In this model we obtain
\begin{eqnarray}
g_{h^0H^+H^-} & = & {-m^2_{h^0}\sin^2\beta\over \sqrt 2 v\cos\beta}
- {m^2_{H^\pm}\sqrt 2\cos\beta\over v} \,, \nonumber \\
g_{H^0H^+H^-} & = & {m^2_{h^0}\cos^2\beta\over \sqrt 2 v\sin\beta}
+ {m^2_{H^\pm}\sqrt 2\sin\beta\over v} \,, \\
g_{AH^+H^-} & = & 0 \,, \nonumber
\end{eqnarray}
where for numerical purposes we have set $\lambda_3$=0 and $\alpha=-\pi/2$.
Since the $AH^+H^-$ coupling vanishes, the $A\to\gamma\gamma$ process can
proceed only via fermion loops. For $m_{h^0}$=60 GeV, we find
$\Gamma(h^0\to\gamma\gamma)$=1.98 MeV.

For substantial production of $h^0$  in association with a $Z^*$ at LEP via
$\epem\to
Z^*h^0$, the $ZZ^*h^0$ coupling ($\sim\sin(\beta-\alpha$)) must
be as large as possible.  With $\alpha$ very close to $-\pi/2$, we note
that $\sin(\beta-\alpha)\simeq 1-\half\beta^2+{\cal O}(\beta^4)$ if $\beta
\ll 1$, and thus we will require $\beta$ to be small.
For $\sin(\beta-\alpha)\simeq 1$, the branching fraction for
$Z\to Z^*h^0\to e^+e^-h^0$ is $0.36\times 10^{-6}$ for $m_{h^0}= 60
\gev$. [L3 has a total of $\sim 10^6$ $Z$ events.] However,
$\beta$ cannot be too small
if the couplings of $H^\pm,\ H^0$, and $A$ are to remain
perturbative and be consistent with low-energy data.
For $\mt=150\gev$, such arguments\cite{bhp} indicate that
$\tan\beta\gsim 1/4$. Henceforth we restrict $\beta$ to lie in
the relatively narrow region $1/5\lsim\beta\lsim 1/2$ to give SM-like coupling
strength to $ZZ^*h^0$ and still satisfy the perturbative
constraints.  For $\beta$ in this range and $\mt=150\gev$, the current
experimental bound\cite{cleo} on the decay $b\to s\gamma$ places a reasonably
strong
constraint\cite{bhp,jlh} on $m_{H^\pm}$.  We obtain $m_{H^\pm}\gsim 45\
(250,\ 420)\gev$ for $\tan\beta=1/2\ (1/3,\ 1/4)$, which suggests that $H^\pm$
may be quite heavy and then $t\to H^+b$ decay would not occur.

To ascertain that the
$h^0\to\gamma\gamma$ mode dominates $h^0$ decay, we must also examine the rates
for such modes as $h^0\to W^*W^*$, $Z^*Z^*$ which have so far only been
calculated
within the SM context{\cite {han}}. With $\alpha=-\pi/2$, $m_{h^0}=$ 60 GeV,
and $\tan \beta$ in the range above we find that the partial widths for these
two process is smaller than that for the $\gamma\gamma$ mode by factors of
order 20--100.

If $Z\to Z^*h^0\to\ell^+\ell^-\gamma\gamma$ is the source of the
L3 events, other $Z^*$ decay modes should also be observed.  L3 has
searched\cite{otherlthree}
for the process $\epem\to Z^*h^0\to q\bar q\gamma\gamma$
and has placed a bound on the cross section of $\simeq 1$ pb for $m_{h^0}\simeq
50$--60\,GeV.  Using the ratio of branching fractions $B(Z\to q\bar q)/
B(Z\to e^+e^-)=20.85$ measured at LEP\cite{sm}, this bound can be
translated into a constraint on the branching fraction for
$Z\to Z^*h^0\to e^+e^-\gamma\gamma$.  The result is
$B\lsim 0.84\times 10^{-6}$, which is not far above the model prediction of
$B=0.36\times10^{-6}$.
Thus a factor of $\sim 2$ increase in statistics
should also reveal the $h^0$ signal in the $q\bar q\gamma\gamma$ mode.
Additionally, since $B(Z\to
\nu\bar\nu)/B(Z\to e^+e^-)\simeq 6$, the LEP experiments should also
observe approximately 3 times as many $2\gamma$ plus missing energy events
as the sum of $e^+e^-\gamma\gamma$ and $\mu^+\mu^-\gamma\gamma$ events.

Since the $W^\mp H^\pm H^0,\ ZH^0A,\ h^0W^+W^-$, and $h^0ZZ$
couplings all
scale as $\sin(\beta-\alpha)$, they will all be large; the $W^\mp H^\pm
h^0,\ Zh^0A,\ H^0W^+W^0$, and $H^0ZZ$ couplings, which scale as $\cos(\beta
-\alpha)$, will all be of ${\cal O}(\beta)$ and will be
suppressed.  This has important
implications for the production of the heavy $H^0$ at the SSC or LHC.
(i)~The
process $gg\to H^0\to W^+W^-$ will be modified in two ways relative to that of
the SM.  First, since the $ggH^0$ coupling is induced by fermion loops, it
will be enhanced by a factor of $\sin\alpha/\sin\beta$.  Second, the $H^0
W^+W^-$ coupling is suppressed by a factor of $\cos(\beta-\alpha)$.
With $\alpha\simeq -\pi/2$ and $1/5\lsim\beta\lsim 1/2$, the
resulting production cross section is essentially that of the SM Higgs boson.
(ii)~The rate
for $W^+W^-\to H^0\to W^+W^-$ is suppressed by a factor of $\cos^4(\beta-
\alpha)\simeq\beta^4$, making this process unimportant for the
production
of a heavy $H^0$.
(iii) For $H^0$'s with intermediate mass, the $WW^*$ decay
mode is suppressed in comparison to the SM, but the $H^0\to\gamma\gamma$ decay
is enhanced.  Hence the number of $H^0\to\gamma\gamma$ events for $H^0$
in this mass range will be larger than in the SM case.  For example, $R_H=
\sigma(gg\to H^0\to\gamma\gamma)/\sigma(gg\to H^0_{SM}\to\gamma\gamma)=57\
(19,\ 18)$ for $m_{H^0}=90\ (120,\ 150)\gev$. A similar, but somewhat smaller
result, $R_A$, is obtained for the corresponding
$gg\to A\to\gamma\gamma$ process.

Figure~1 shows the ratios $R_{H,A}$ as
functions of the respective particle masses; note that $R_A$ is always
less than $\simeq$ 1 for light $A$ but grows rapidly for $m_A$ above 140 GeV.
(iv) The rate for the process $gg\to H\to t\bar t$
is found to be enhanced in comparison to the SM by a factor of
$\sin^{-4}\beta \simeq$ 100 which may render it observable at the SSC/LHC.

Figures 2a and 2b show the $H^0$ and $A$ branching fractions, respectively,
for $\mt=150\gev,\
\tan\beta=1/3,\ \alpha=-\pi/2$, and $m_{H^\pm}=600\gev$.  For this choice
of parameters, the full widths of both $H^0$ and $A$ are
comparable to that of the SM Higgs boson. Figure 2c compares
these full widths over a wide range of masses.  For a
sufficiently massive $H^0$ or $A$, the decay into
the other physical Higgs fields (\ie, $H^0\to H^+H^-,\ 2A$, \etc.) will be
dominant. Note that couplings such as $W^\mp H^\pm A$
and $ZH^+H^-$
are fixed by gauge invariance and are unaffected by our choice of the
parameters $\alpha$ and $\beta$.

In summary, a two-Higgs doublet model in which one doublet primarily
generates the $W, Z$ masses and the other doublet is dominately coupled
to fermions has interesting phenomenological implications which deserve
further attention.  Such a model could account for the $\epem\gamma\gamma$
and $\mu^+\mu^-\gamma\gamma$ events observed by the L3 Collaboration.  If
this interpretation is correct, $\gamma\gamma +$ missing energy and
$\gamma\gamma +$ two-jet events should also be found at LEP experiments.

\vskip.25in
\centerline{ACKNOWLEDGEMENTS}

The authors would like to thank R.J.N.\ Phillips for discussions related
to this work.
J.L.H. and T.G.R. wish to thank the Institute of Theoretical Science at the
University of Oregon for
their hospitality during the course of this work.
This work was supported in
part by the U.S.~Department of Energy
under contracts W-31-109-ENG-38, DE-FG-85ER400224, and DE-AC02-76ER00881,
and in part Texas National Laboratory Commission under grant no. RGFY9173,
and in part by the University of Wisconsin Research Committee with funds
granted by the Wisconsin Alumni Research Foundation.

\newpage

%
\def\MPL #1 #2 #3 {Mod.~Phys.~Lett.~{\bf#1},\ #2 (#3)}
\def\NPB #1 #2 #3 {Nucl.~Phys.~{\bf#1},\ #2 (#3)}
\def\PLB #1 #2 #3 {Phys.~Lett.~{\bf#1},\ #2 (#3)}
\def\PR #1 #2 #3 {Phys.~Rep.~{\bf#1},\ #2 (#3)}
\def\PRD #1 #2 #3 {Phys.~Rev.~{\bf#1},\ #2 (#3)}
\def\PRL #1 #2 #3 {Phys.~Rev.~Lett.~{\bf#1},\ #2 (#3)}
\def\RMP #1 #2 #3 {Rev.~Mod.~Phys.~{\bf#1},\ #2 (#3)}
\def\ZP #1 #2 #3 {Z.~Phys.~{\bf#1},\ #2 (#3)}

\newpage

%
{\bf Figure Captions}
\begin{itemize}

\item[Figure 1.]{The ratio of cross sections, $R_{H,A}=\sigma(gg\to H^0,A\to
\gamma\gamma)/\sigma(gg\to H_{SM}\to \gamma\gamma)$, as a function of
the appropriate Higgs mass.}
\item[Figure 2.]{The branching fractions, B, for the decay of (a) $H^0$ and
(b) $A$ versus mass, and (c) the total widths as functions of the
Higgs mass for $H^0$ and
$A$ compared to the SM Higgs width.}
\end{itemize}

\end{document}